\RequirePackage{amsmath}
\documentclass[runningheads]{llncs}
\usepackage[utf8]{inputenc}
\usepackage{colortbl}
\usepackage{amssymb}
\usepackage{amsbsy}
\usepackage{hyperref}
\usepackage{float}
\usepackage{stmaryrd}
\usepackage{subcaption}
\usepackage[font=scriptsize]{caption}
\usepackage{mathtools}
\usepackage[ruled,vlined,linesnumbered]{algorithm2e}
\usepackage{tabularx}
\usepackage{media9}
\usepackage{xcolor}
\usepackage{booktabs} 
\usepackage{tikz}
\usepackage[xindy]{glossaries}

\newcommand{\Gwen}{\textsc{GWENDOLEN}}

\newcommand{\eventually}{\ensuremath{\lozenge}}
\newcommand{\lbelief}[2]{\ensuremath{\mathrm{{\cal B}}_{#1} \, #2}}

\newcommand{\always}{\ensuremath{\Box}}

\newacronym{rml}{RML}{Runtime Monitoring Language}

\newacronym{ltl}{LTL}{Linear-time Temporal Logic}

\newacronym{ctl}{CTL}{Computation Tree Logic}

\newacronym{ptl}{PTL}{Probabilistic Temporal Logic}

\newacronym{pctl}{PCTL}{Probabilistic Computation Tree Logic}

\newacronym{pctl*}{PCTL*}{Probabilistic Computation Tree Logic*}

\newacronym{tctl}{TCTL}{Timed Computation Tree Logic}

\newacronym{tltl}{TLTL}{Timed Linear-time Temporal Logic}

\newacronym{rvltl}{RV-LTL}{Runtime Verification Linear Temporal Logic}

\newacronym{stl}{STL}{Signal Temporal Logic}

\newacronym{ltl-x}{LTL-X}{a version of Linear Time Logic without the `next' operator}

\newacronym{mtl}{MTL}{Metric Temporal Logic}

\newacronym{iactlk-x}{IACTLK-X}{a fragment of the temporal-epistemic logic CTLK, without the `next' operator}

\newacronym{fotl}{FOTL}{First-Order Temporal Logic}

\newacronym{bdi}{BDI}{Belief-Desire-Intention}

\newacronym{prs}{PRS}{Procedural Reasoning System}

\newacronym{are}{ARE}{Autonomous Reasoning Engine}

\newacronym{dmars}{dMARS}{distributed Multi-Agent Reasoning System}

\newacronym{fsm}{FSM}{Finite-State Machine}

\newacronym{pfsm}{PFSM}{Probabilistic Finite-State Machine}

\newacronym{apfsm}{APFSM}{\textit{Autonomous} Probabilistic Finite-State Machine}

\newacronym{pha}{PHA}{Probabilistic Hybrid Automata}

\newacronym{fsa}{FSA}{Finite-State Automata}

\newacronym{ta}{TA}{Timed Automata}

\newacronym{gspn}{GSPN}{Generalised Stochastic Petri Net}

\newacronym{mopn}{MOPN}{Marked Ordinary Petri Net}

\newacronym{ba}{BA}{B\"{u}chi Automata}

\newacronym{mdp}{MDP}{Markov Decision Process}

\newacronym{dtmc}{DTMC}{Discrete-Time Markov Chain}

\newacronym{pars}{PARS}{Process Algebra for Robot Schemas}

\newacronym{fsp}{FSP}{Finite State Processes}

\newacronym{piadl}{$\pi$ADL}{$\pi$ calculus combined with ADL}

\newacronym{csp}{CSP}{Communicating Sequential Processes}

\newacronym{ccs}{CCS}{Calculus of Communicating Systems}

\newacronym{wsccs}{WSCCS}{Weighted Synchronous CCS}

\newacronym{csp|b}{CSP$\|$B}{an integrated formalism combining CSP and B}

\newacronym{fdr}{FDR}{the Failures-Divergences Refinement checker}

\newacronym{jpf}{JPF}{Java PathFinder}

\newacronym{ajpf}{AJPF}{Agent Java PathFinder}

\newacronym{mcmas}{MCMAS}{Model Checker for Multi-Agent Systems}

\newacronym{ltlmop}{LTLMoP}{Linear Temporal Logic MissiOn Planning}

\newacronym{adl}{ADL}{Architecture Description Language}

\newacronym{aadl}{AADL}{Architecture Analysis and Design Language}

\newacronym{lts}{LTS}{Labelled-Transition System}

\newacronym{uml}{UML}{Unified Modelling Language}

\newacronym{sysml}{SySML}{Systems Modelling Language}

\newacronym{robotml}{RobotML}{Robot Modelling Language}

\newacronym{dsl}{DSL}{Domain Specific Language}

\newacronym{sct}{SCT}{Supervisory Control Theory}

\newacronym{ide}{IDE}{Integrated Development Environment}

\newacronym{mde}{MDE}{Model-Driven Engineering}

\newacronym{ai}{AI}{Artificial Intelligence}

\newacronym{fdir}{FDIR}{Fault Detection, Isolation and Recovery}

\newacronym{ifm}{iFM}{Integrated Formal Methods}

\newacronym{ros}{ROS}{Robot Operating System}

\newacronym{dl}{\text{\upshape\textsf{d{\kern-0.05em}L}}}{differential dynamic logic}

\newacronym{vm}{VM}{Virtual Machine}

\newacronym{bip}{BIP}{Behaviour Interaction Priority}

\newacronym{fol}{FOL}{First-Order Logic}

\newacronym{owl}{OWL}{Web Ontology Language}

\newacronym{ria}{RIA}{Restore Invariant Approach}

\newacronym{ocl}{OCL}{Object Constraint Language}

\newacronym{sil}{SIL}{Safety Integrity Level}

\newacronym{rule}{RULE}{Robotic Urban-Like Environment}

\newacronym{forms}{FORMS}{FOrmal Reference Model for Self-adaptation}

\newacronym{rv}{RV}{Runtime Verification}

\newacronym{prv}{PRV}{Predictive Runtime Verification}

\newacronym{mmprv}{MMPRV}{Multi-Model Predictive Runtime Verification}

\newacronym{cmmprv}{CMMPRV}{Centralised Multi-Model Predictive Runtime Verification}

\newacronym{dmmprv}{DMMPRV}{Decntralised Multi-Model Predictive Runtime Verification}

\newacronym{sua}{SUA}{System Under Analysis}

\newacronym{race}{RACE}{Remote Applications in Challenging Environments}

\newacronym{ccfe}{CCFE}{Culham Centre for Fusion Energy}\RequirePackage{amsmath}

\usetikzlibrary{shapes.geometric, arrows}
\tikzstyle{process} = [rectangle, minimum width=3em, minimum height=2em, text centered, draw=black]
\tikzstyle{arrow} = [thick,->,>=stealth]
\usetikzlibrary{shapes,backgrounds,arrows,calc,positioning,snakes}

\pdfpageattr {/Group << /S /Transparency /I true /CS /DeviceRGB>>}
\usepackage{adjustbox}
\usepackage{xspace}
\usepackage{zed-csp}
\usepackage{gensymb}
\usepackage{lscape}
\usepackage[misc,geometry]{ifsym}

\SetKwProg{Fn}{Function}{}{}

\usepackage{color}

\usepackage[textsize=normalsize,textwidth=3cm]{todonotes}

\begin{document}
\title{Heterogeneous Verification of an\\ Autonomous Curiosity Rover\thanks{Work supported by UK Research and Innovation, and EPSRC Hubs for ``Robotics and AI in Hazardous Environments'': EP/R026092 (FAIR-SPACE) and EP/R026084 (RAIN).}}
\author{Rafael C. Cardoso$^{\textrm{(\Letter)}}$ \and Marie Farrell \and Matt Luckcuck \and Angelo Ferrando \and Michael Fisher}
\institute{Department of Computer Science, University of Liverpool, Liverpool L69 3BX, UK\\
\email{\{rafael.cardoso,marie.farrell,m.luckcuck,\\angelo.ferrando,mfisher\}@liverpool.ac.uk}}
\authorrunning{R. C. Cardoso et al.}

\maketitle

\begin{abstract}
The Curiosity rover is one of the most complex systems successfully deployed in a planetary exploration mission to date. It was sent by NASA to explore the surface of Mars and to identify potential signs of life. Even though it has limited autonomy on-board, most of its decisions are made by the ground control team. This hinders the speed at which the Curiosity reacts to its environment, due to the communication delays between Earth and Mars. Depending on the orbital position of both planets, it can take 4--24 minutes for a message to be transmitted between Earth and Mars. If the Curiosity were controlled autonomously, it would be able to perform its activities much faster and more flexibly. However, one of the major barriers to increased use of autonomy in such scenarios is the lack of assurances that the autonomous behaviour will work as expected. In this paper, we use a Robot Operating System (ROS) model of the Curiosity that is simulated in Gazebo and add an autonomous agent that is responsible for high-level decision-making. Then, we use a mixture of formal and non-formal techniques to verify the distinct system components (ROS nodes). This use of heterogeneous verification techniques is essential to provide guarantees about the nodes at different abstraction levels, and allows us to bring together relevant verification evidence to provide overall assurance.
\end{abstract}

\section{Introduction}


We present a case study with a simulation of the Curiosity rover undertaking an exploration mission. Crucially, we have equipped the rover with decision-making capabilities so that it does not rely on human teleoperation. As a result of the added autonomous behaviour, it is important to provide safety assurances about critical components in the system. Usually, components in such systems are modular and each individual component often requires a different verification technique(s)~\cite{webster2016corroborative,luckcuck2019formal,Farrell2018}. We have applied distinct verification techniques to various critical components and at different abstraction levels to ensure the correctness of the overall system. All of the artefacts (source code, videos, etc.) discussed in this paper are available in our online repository\footnote{\url{https://github.com/autonomy-and-verification-uol/curiosity-NFM2020}}.

\section{Mission Description, Simulation and Autonomy}
\label{sec:mission}
\paragraph{Mission Description:} We simulate an inspection mission, where the Curiosity patrols a topological map of the surface of Mars. We assume that the map is known prior to this mission, and in this paper we only consider a small subset of the map (i.e. the agent has map coordinates for each waypoint in the map). Specifically, we consider four different waypoints ($o$, $A$, $B$, and $C$) that are spread across the Martian terrain. Low-level movement is achieved through a dead reckoning or feedback control.

\begin{figure}[t]
\centering
\scalebox{0.7}{
\begin{tikzpicture}[node distance=2em]
  \node (origin) [draw, circle] {$o$};
  \node (A) [draw,circle,right of = origin, xshift = 5em, fill=gray!30] {$A$};
  \node (B) [draw,circle,right of = origin, xshift = 5em, yshift = -4em,fill=yellow!50] {$B$};
  \node (C) [draw,circle,right of = origin, xshift = 5em, yshift = -8em] {$C$};
  
  \node(windy) [draw,circle,right of = A, xshift = 5em, fill=gray!30] {};
  \node(windytext)[right of = windy, xshift = 0.5em]{Windy};
  
  \node(rad) [draw,circle,below of = windy, fill=yellow!50] {};
  \node(radtext)[right of = rad, xshift = 1.1em]{Radiation};
  
  \draw [arrow] (origin) -- node[below]{}(A);
  \draw [arrow] (A) -- node[below]{}(B);
  \draw [arrow] (B) -- node[below]{}(C);
  \draw [arrow] (C) to [out=0,in=0] node[below]{}(A);
  
\end{tikzpicture}}
    \caption{The Curiosity  begins at the origin, $o$, and then visits the waypoints $A$, $B$ and $C$ in whichever order is safe. We indicate waypoints with high levels of wind (grey) and radiation (yellow).}
    \label{fig:waypoints}
\end{figure}
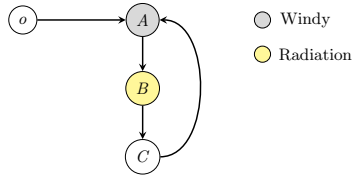

We begin with the deployment of the Curiosity and a startup period where it initialises all three of its control modules (wheels, arms, and mast). After the agent receives confirmation that the modules are ready, it autonomously controls the Curiosity to move between the waypoints in the following order: ($o \rightarrow A \rightarrow B \rightarrow C \rightarrow A \rightarrow \ldots$), as shown in Fig.~\ref{fig:waypoints}. This is the ideal scenario, however, if one of the waypoints is experiencing high levels of radiation then the rover should skip it until the radiation has reduced to a safe level. For data collection, the mast and arm should be open, however, it is unsafe to do so in windy conditions. We do not model battery power. Instead, we assume that the rover has sufficient battery power to traverse the waypoints and operate the equipment.


\paragraph{Simulation:}

We obtained a \gls{ros}~\cite{quigley2009ros} version of the Curiosity from a \gls{ros} teaching website\footnote{\url{https://bitbucket.org/theconstructcore/curiosity_mars_rover/src/master/}} which uses official data and 3D models of the Curiosity and Martian terrain which have been made public by NASA. This \gls{ros} simulation runs in Gazebo\footnote{\url{http://gazebosim.org/}}, a 3D simulator. 
Most of the Curiosity's effectors are included in the simulation. 
It has the complete chassis of the rover with all six wheels and the suspension system, a retractable arm with four joints, and a retractable mast with two joints and a camera (Mastcam) on top. Some of the sensors are missing, e.g. MAHLI (Mars Hand Lens Imager), as these would require simulated sensor data.


In the original configuration, the standard control method of the Curiosity was implemented using \gls{ros} services and it was controlled via teleoperation. \gls{ros} services are defined as a pair of request and reply messages that are provided by \gls{ros} nodes. 
In our simulation, we re-implemented the control method through action libraries, which follow a client-server model that is similar to \gls{ros} services. Both can receive a request to perform some task and then generate a reply. The difference in using action libraries is that the client can cancel the action, as well as receive feedback about the task execution. Thus, action libraries are more suited for use with decision-making agents since they allow more fine-grained control.

We developed three action libraries: one each for the wheels, arm, and mast. The wheels client receives high-level action commands to move forward, backward, left, and right; or a waypoint from the topological map (using the move base library for path planning). Based on the command received, the server controls each of the six wheels and publishes speed commands to the appropriate wheels depending on the direction or topological waypoint requested in the action. If a direction command is given, then the server expects three parameters: direction of movement, speed, and distance. After a movement action, the server always calls a stop action that sets the speed of all wheels to zero. The arm and mast action libraries control the joints of their respective effectors so that they can be positioned correctly for use.

\paragraph{Enabling Autonomous Decision-Making:}
We use the \Gwen{}~\cite{Dennis08} agent programming language to implement the high-level control and autonomous decision-making behaviour of the Curiosity. Agent programming languages abstract the environment and other external sources, focusing on high-level autonomous control, resulting in smaller and more modular code than other languages. Due to the agent's reasoning cycle an execution trace can clearly show how the agent came to a decision, thus providing us with explainability. Using \Gwen{} allows us to verify properties of the agent's reasoning, allowing the safeguard of critical behaviours.

\Gwen{} agents follow the \gls{bdi} model~\cite{rao:95b}. Beliefs, desires, and intentions represent respectively the information, motivational, and deliberative states of the agent. We developed a \Gwen{} environment that communicates with \gls{ros} through the \emph{rosbridge} library. 
When the agent executes an action in the environment, the action is processed and published to the action's associated \gls{ros} topic. The environment creates subscribers that listen to specific topics so that necessary perceptions are created and sent to the agent.

In the Curiosity simulation, the \Gwen{} agent has four high-level actions. The action \emph{control\_wheels} has three parameters: direction of movement (forward, backward, left, or right), speed (an integer with sign to indicate direction), and distance (in seconds). The \emph{move\_to\_waypoint} action contains one parameter with a waypoint from the topological map. The actions, \emph{control\_arm} and \emph{control\_mast}, both have one parameter whose possible values are either \emph{open} or \emph{close}.

Fig.~\ref{fig:system} illustrates a high-level system diagram with the communication paths between the nodes in the simulation. We have verified distinct components of this simulation using different methods. Specifically, we verify the autonomous agent using the AJPF program model checker; the interface that this agent has with the environment using Dafny; a CSP specification of the action library nodes using FDR4; and we use each of these formal models to guide the generation of runtime monitors. This combination of simulation-based testing, and the use of multiple formal methods at different levels of abstraction, gives us a basis for providing assurances about the use of autonomous decision-making in this extreme environment mission scenario, and could be transferred and applied to other similar case studies as shown in~\cite{webster2016corroborative,Farrell2018,luckcuck2019formal}.

\begin{center}
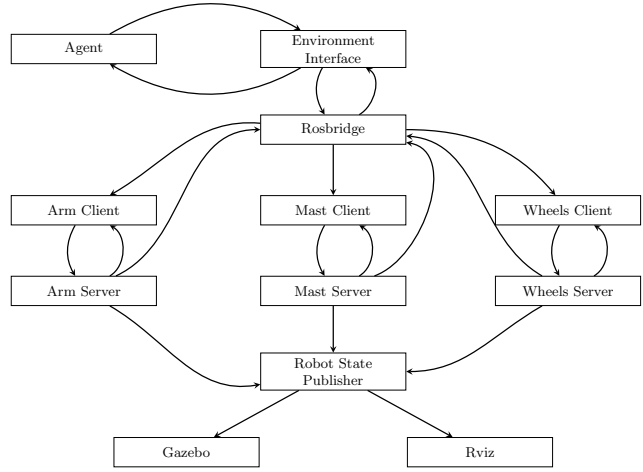
\begin{figure*}[t]
\centering
\scalebox{0.6}{
\begin{tikzpicture}[node distance=1.5em]

\node (env) [process] {\parbox[t][][t]{3cm}{\centering Environment Interface}};

\node (agent) [process, xshift = -17em] {\parbox[t][][t]{3cm}{\centering Agent}};

\node (rosbridge) [process, below of = env, yshift = -4em] {\parbox[t][][t]{3cm}{\centering Rosbridge}};

\node (armclient) [process, below of = rosbridge, yshift = -4em, xshift = -17em] {\parbox[t][][t]{3cm}{\centering Arm Client}};

\node (mastclient) [process, below of = rosbridge, yshift = -4em] {\parbox[t][][t]{3cm}{\centering Mast Client}};

\node (wheelsclient) [process, below of = rosbridge, yshift = -4em, xshift = 16em] {\parbox[t][][t]{3cm}{\centering Wheels Client}};

\node (armserver) [process, below of = armclient, yshift = -4em] {\parbox[t][][t]{3cm}{\centering Arm Server}};

\node (mastserver) [process, below of = mastclient, yshift = -4em] {\parbox[t][][t]{3cm}{\centering Mast Server}};

\node (wheelsserver) [process, below of = wheelsclient, yshift = -4em] {\parbox[t][][t]{3cm}{\centering Wheels Server}};

\node (robotstate) [process, below of = mastserver, yshift = -4em] {\parbox[t][][t]{3cm}{\centering Robot State Publisher}};

\node (gazebo) [process, below of = robotstate, yshift = -4em, xshift = -10em] {\parbox[t][][t]{3cm}{\centering Gazebo}};

\node (rviz) [process, below of = robotstate, yshift = -4em, xshift = 10em] {\parbox[t][][t]{3cm}{\centering Rviz}};

\draw [arrow] (env) to [out=-120,in= 120] node[right]{}(rosbridge);
\draw [arrow] (agent) to [out=30,in=150] node[right]{}(env);
\draw [arrow] (env) to [out=210,in=-30] node[right]{}(agent);
\draw [arrow] (rosbridge) to [out=30,in=-30] node[right]{}(env);
\draw [arrow] (rosbridge) to [out=175,in=30] node[right]{}(armclient);
\draw [arrow] (rosbridge) -- node[right]{}(mastclient);
\draw [arrow] (rosbridge) to [out=0,in=130] node[right]{}(wheelsclient);

\draw [arrow] (armclient) to [out=-120,in=120] node[left]{}(armserver);
\draw [arrow] (mastclient) to [out=-120,in= 120] node[right]{}(mastserver);
\draw [arrow] (wheelsclient)to [out=-120,in= 120] node[left]{}(wheelsserver);
\draw [arrow] (armserver) to [out=30,in=-30] node[left]{}(armclient);
\draw [arrow] (mastserver) to [out=30,in=-30] node[left]{}(mastclient);
\draw [arrow] (wheelsserver) to [out=30,in=-30] node[right]{}(wheelsclient);

\draw [arrow] (armserver) to [out=25,in=180] node[left]{}(rosbridge);
\draw [arrow] (mastserver) to [out=20,in=-10] node[right, at start, yshift =10pt, xshift =15pt ]{}(rosbridge);
\draw [arrow] (wheelsserver) to [out=150,in=-5] node[right]{}(rosbridge);

\draw [arrow] (armserver) to [out=-30,in=190] node[below]{}(robotstate);
\draw [arrow] (mastserver) -- node[right]{}(robotstate);
\draw [arrow] (wheelsserver) to [out=210,in=0] node[left]{}(robotstate);.

\draw [arrow] (robotstate) -- node[below]{}(gazebo);
\draw [arrow] (robotstate) -- node[below]{}(rviz);

\end{tikzpicture}}

\caption{Overview of the system. Arrows indicate data flow between the nodes.}
\label{fig:system}
\vspace{-10pt}
\end{figure*}
\end{center}

\section{Verification}
This section describes our verification of four critical areas of our simulation of an autonomous Curiosity rover. We verify properties of this system at different levels of abstraction. We begin by describing how we verify that the agent, which is fundamentally controlling the system, makes the correct decisions about which waypoint to visit next.

Next, we discuss our use of an automated theorem prover to verify that the information that the agent receives from the environment sensors is interpreted and acted upon correctly. Then, we outline how we verified that the communication between the \emph{client} and \emph{server} action library nodes (as shown in Fig.~\ref{fig:system}) functions correctly. Finally, we outline our \gls{rv} of design-time assumptions about the environment. Interestingly, we used the preceding formal models as a way to focus these runtime checks on appropriate properties.

\paragraph{Verifying the Agent using AJPF: }
\label{sec:agent}
\label{sec:agentGwen}

Model-checking~\cite{Clarke00:MC} exhaustively examines the state space to check if some desired property holds. This can be applied to either a formal model of the system, encoded in some specification language, or directly to the implementation. The property to be verified is usually specified in a logic-based language. For example, we may want to verify that the Curiosity will not move its arm while collecting soil and rock data, in order to protect the sample.

\gls{ajpf}~\cite{Dennis2012}, an extension of \gls{jpf}~\cite{Visser2003}, is a model-checker that works directly on Java program code. This extension facilitates formal verification of \gls{bdi}-based agent programs by providing a property specification language based on \gls{ltl} that supports the description of terms usually found in \gls{bdi} agents.

For example, some of the properties that we verified of the implementation of our agent were as follows:\\
\centerline{$\always(A_\texttt{rover} move\_to\_waypoint(A) \rightarrow \eventually \lbelief{\texttt{rover}} (at(A)))$}
\centerline{$\always(A_\texttt{rover} move\_to\_waypoint(B) \rightarrow \eventually \lbelief{\texttt{rover}} (at(B)))$}
\centerline{$\always(A_\texttt{rover} move\_to\_waypoint(C) \rightarrow \eventually \lbelief{\texttt{rover}} (at(C)))$}

These properties state that it is always the case ($\always$) that if the \emph{rover} agent executes the action \emph{move\_to\_waypoint} (to either A, B, or C), then eventually ($\eventually$) the \emph{rover} agent will believe that it is currently located in that waypoint.

The syntax of the \gls{ajpf} specification language is limited to expressing agent related properties, such as beliefs, goals, actions, and intentions of a specific agent that was written in \Gwen{}. Moreover, properties specified in \gls{ajpf} must be ground (i.e. cannot be parameterised). For verifying the interface between the agent and the environment, we employ the Dafny program verifier.

\paragraph{Verifying the Agent-Environment Interface: }
\label{sec:agentDafny}
Dafny facilitates the use of specification constructs e.g. pre-/post-conditions, loop invariants and variants~\cite{leino2010dafny}. Dafny is used in the static verification of functional program correctness. Programs are translated into the Boogie intermediate verification language~\cite{barnett2005boogie} and then the Z3 automated theorem prover discharges the associated proof obligations~\cite{de2008z3}. 

Our Dafny model centres on the decisions made by the agent in response to the input that it receives from the environment. In this simple model, we verify an important safety property that the rover will not select any actions if the arm, mast or wheels have not been initialised yet. This is specified as follows:

\centerline{\small{\texttt{\textbf{ensures} (wheelsready \&\& armready \&\& mastready) == false}\texttt{ ==> actions ==[];}}}

\noindent Here, \texttt{wheelsready}, \texttt{armready} and \texttt{mastready} are boolean flags that are toggled by the associated modules, and \texttt{actions} is the sequence of returned actions.

Our Dafny model has functions for accessing the environmental conditions at a given waypoint e.g. \texttt{getEnvironment()} and \texttt{getWind()}. This allows us to verify properties about the how the environmental conditions affect where the rover goes. 
 The \texttt{getEnvironment()} method then checks the wind and radiation at a particular waypoint and we verify that the following condition is met where \texttt{e} is a variable that represents the current status of the environment:

\centerline{\small{\texttt{ensures windspeed < 5 \&\& radiation < 5 ==> e == Fine;}}}

\noindent In this way, our Dafny model allows us to verify conditions about the safety of the agent and also that the information coming from the environment is interpreted correctly by the agent. We provide other verified methods including \texttt{getRad()} which is a high-level implementation of how the radiation at waypoint $B$ decays over time. Our loop invariant in the \texttt{CuriosityAgent()} also ensures that the rover can't be at waypoint $B$ when the radiation is too high:

\centerline{\small{\texttt{invariant !(current == B  \&\& env == Radiation);}}}

We included radiation at $B$ in the Dafny implementation to examine how the rover reacts to radiation at a particular waypoint, as per the mission description (Sect. \ref{sec:mission}). Next, we verify the action library client and server nodes using CSP.

\paragraph{Verifying Action Library Communication:}
\label{sec:nodeCSP}

We verify the communication between the pairs of action library client--server nodes that interface between the software and hardware (arm, mast, and wheels). Each client accepts instructions from the agent (via \emph{rosbridge}) which it then sends to the relevant server node as a goal (task to complete). Since the \gls{ajpf} model checker can only check agent-programs, we decided to use \gls{csp} to verify this critical link. \gls{csp} processes describe sequences of events; $a \then b \then \Skip$ is the process where events $a$ and $b$ occur sequentially, then terminates ($\Skip$). 

The \gls{csp} model is constructed from the Curiosity \gls{ros} code, capturing both the program-specific and the generic action library behaviour. Each of the client--server pairs is modelled by one \gls{csp} file, with one further file modelling the generic behaviour of an action library server. We use the FDR4 model-checker~\cite{GibsonRobinson2014} to check three properties: (1) when a client sends a goal, it will begin execution on the correct server, (2) when a client sends a goal, eventually it receives a result from the server, and (3) when the agent instructs a client node to perform an action, the server informs the agent that it is ready and then eventually the agent receives a result. Here we give an example of (2), where we check that if the arm client sends a goal then eventually it will receive a result:\\
\centerline{$send\_goal\_arm?\_ \rightarrow executeGoal.arm \rightarrow SKIP$}

\paragraph{Runtime Verification:}
\label{sec:rv}
It is achieved by examining the current execution of the system at runtime against a formal specification. Since runtime monitors only observe the current system execution, the resulting approach is not exhaustive in the sense that model-checking is (which examines the entire state space). 
However, monitor implementations are usually extremely efficient since they do not consider all possible system executions and they can remain as safeguards after deployment. In this way, a monitor helps to ensure correct system behaviour.

ROSMonitoring\footnote{\url{https://github.com/autonomy-and-verification-uol/ROSMonitoring}} (ROSMon) is a flexible and formalism-agnostic \gls{rv} framework for ROS. 
ROSMon creates gaps in the communication between nodes in the system. 
These gaps are then filled by monitors which are automatically synthesised by ROSMon. In this way, the messages of interest are forced to pass through the monitors and are checked against a corresponding formal specification. 
We applied ROSMon to our simulation to check properties at runtime. 
For example, using Dafny, we verify the agent-environment interface; ROSMon bridges the gap between the Dafny model and the real environment by checking at runtime if the assumptions used in the Dafny model are satisfied by the real system.

We used a property, written in \gls{rml}, to synthesise a monitor to check the constraint used in the Dafny \texttt{getEnvironment} method. Here, we check that the wind speed and radiation are always positive, and if the wind speed and radiation are less than 5 each, then the environment is ``Fine''. This is (partially) written as follows:

{\fontsize{8}{8}\selectfont
\begin{verbatim}
Main = (GetEnvironmentConstraints /\ (wind_speed(_) >> wind_speed_at_least(0)*) /\
        (radiation_units(_) >> radiation_units_at_least(0)*));
GetEnvironmentConstraints =
  wind_speed_up_to(4) GetEnvironmentConstraints1 
  \/ radiation_units_up_to(4) GetEnvironmentConstraints2 
  \/ any GetEnvironmentConstraints;
...
\end{verbatim}
}

In this way, we used abstract formal system models to guide the development of corresponding runtime monitors to examine these properties at runtime.

\section{Discussion}
\label{sec:conclusion}
This paper has reported on our case study of using multiple verification techniques to provide assurance for an autonomous Curiosity rover undertaking an exploration/sampling mission. We used the \Gwen{} agent programming language to implement an autonomous agent in a \gls{ros}-based simulation of the Curiosity. We verified this agent using \gls{ajpf}, how it responds to discrete input from its environment using Dafny, the message passing between the action library nodes using CSP, and we synthesised runtime monitors using ROSMon. 

We employed a myriad of verification techniques to verify the behaviour of distinct aspect(s) of the system. Our aim was to streamline the process of verifying the system by verifying each system component using a suitable technique, rather than attempting to verify everything using only one technique. For example, we use an agent programming language for the agent and CSP for message passing. The tool used to verify the agent program is not appropriate (and would generally not work) to verify message passing.

Our use of RV is of particular interest here since the system is implemented in C++ or Python for which formal verification at code level is not currently feasible/possible. However, the tools and techniques that were chosen are not necessarily the only ones that were suitable for any specific component and certainly other choices could have been made. Future work includes investigating these alternatives. 

Our use of heterogeneous techniques for various critical components of the system was motivated by the work done in~\cite{webster2016corroborative,Farrell2018,luckcuck2019formal}. Our case study exhibits how heterogeneous verification techniques can be applied to various components of an autonomous robotic system at different levels of abstraction. Future work seeks to link the results of these heterogeneous techniques in a holistic framework so that they might inform one another.

\clearpage
\bibliographystyle{splncs04}
\bibliography{nfm}

\end{document}